\documentclass{aa}  

\usepackage{graphicx}
\usepackage{txfonts}
\usepackage{bm}
\usepackage{xcolor}

\begin{document}

   \title{Cyanogen, cyanoacetylene, and acetonitrile in comet 67P and their relation to the cyano radical}
   \titlerunning{NCCN, CH$_3$CN, and HC$_3$N in comet 67P and their relation to CN}
        

\author{N. H{\"a}nni\inst{1} \and K. Altwegg\inst{1} \and H. Balsiger\inst{1} \and M. Combi\inst{2} \and S. A. Fuselier\inst{3,4} \and J. De Keyser\inst{5} \and B. Pestoni\inst{1} \and M. Rubin\inst{1} \and S. F. Wampfler\inst{6}}

\institute{Physics Institute, Space Research \& Planetary Sciences, University of Bern, Sidlerstrasse 5, CH-3012 Bern, Switzerland\\ \email{nora.haenni@space.unibe.ch}
\and
Department of Climate and Space Sciences and Engineering, University of Michigan, Ann Arbor, MI, USA.
\and
Space Science Directorate, Southwest Research Institute, San Antonio, TX, USA.
\and
Department of Physics and Astronomy, The University of Texas at San Antonio, San Antonio, TX, USA.
\and
Royal Belgian Institute for Space Aeronomy, BIRA-IASB, Brussels, Belgium.
\and
Center for Space and Habitability, University of Bern, Gesellschaftsstrasse 6, CH-3012 Bern, Switzerland}

\date{Received XXX / Accepted YYY}


\abstract
{The cyano radical (CN) is one of the most frequently remotely observed species in space, and is also often observed in comets. Data for   the inner coma of comet 67P/Churyumov-Gerasimenko collected by the high-resolution
Double Focusing Mass Spectrometer (DFMS) on board the Rosetta orbiter revealed an unexpected chemical complexity, and, recently, also more CN than expected from photodissociation of its most likely parent, hydrogen cyanide (HCN). Here, we derive abundances relative to HCN of three cometary nitriles (including structural isomers) from DFMS data. Mass spectrometry of complex mixtures does not always allow isolation of structural isomers, and therefore in our analysis we assume the most stable and abundant (in similar environments) structure, that is HCN for CHN, CH$_3$CN for C$_2$H$_3$N, HC$_3$N for C$_3$HN, and NCCN for C$_2$N$_2$. For cyanoacetylene (HC$_3$N) and acetonitrile (CH$_3$CN), the complete mission time-line was evaluated, while cyanogen (NCCN) 
was often below the detection limit. By carefully selecting periods where cyanogen was above the detection limit, we were able to follow the abundance ratio between NCCN and HCN from 3.16 au inbound to 3.42 au outbound. These are the first measurements of NCCN in a comet. We find that neither NCCN nor either of the other two nitriles is sufficiently abundant to be a relevant alternative parent to CN.}


\maketitle

\section{Introduction}\label{sec:intro}
The CN radical is thought to be a key-species in the evolution of prebiotic molecules and its abundance in various stellar and interstellar environments is well established from remote observations \citep{herbst2009,vandishoeck2014}. \citet{haenni2020} reported the first in situ detection of the CN radical in comet 67P/Churyumov-Gerasimenko (67P hereafter). These latter authors showed that the abundance of CN cannot be explained by photodissociation of HCN alone. These results are consistent with various previous remote observation studies of other comets, for example of 6P/d'Arrest observed by \citet{dellorusso2009}. The first comet showing this discrepancy was C/1983 H1 (IRAS-Araki-Alcock), observed by \citet{ahearn1983,bockelee-morvan1984}. Haser model \citep[cf.][]{haser1957} calculations showed that for certain comets, production rates of CN and HCN are not compatible \citep{bockelee-morvan1985,fray2005,dellorusso2016}. Because of the quantum yield of 0.97 of the HCN photodissociation into CN, production rates of the two molecules in comets should be of the same order of magnitude.\footnote{The yield of a daughter species (D) from a parent species (P) is determined by two factors: the photodissociation rate of P (i.e., the number of dissociations per second) and the quantum yield of P $\rightarrow$ D (i.e., the portion of dissociations that occur according to this specific dissociation channel).} Half of Fray's sample of eight comets showed inconsistent production rates and the full sample showed  shorter CN formation scale lengths than the HCN destruction scale length, especially below 3~au. Possible additional or alternative parents of CN should therefore have  higher photodissociation rates than that of HCN, which was estimated by \citet{crovisier1994} to be $1.5 \times 10^{-5}$~s$^{-1}$ and by \citet{huebner1992} to be $1.26 \times 10^{-5}$~s$^{-1}$ at quiet Sun conditions at 1~au. The situation appears slightly different if production rates obtained from infrared (IR) spectroscopy are considered, such as those reviewed by \citet{dellorusso2016}, because the IR-based HCN production rate is significantly higher for some comets than the radio-based one. The origin of the observed discrepancies is the topic of ongoing debate. However,  \citet{dellorusso2016} also reported incompatible CN and HCN production rates for 7 out of the 21 comets investigated. These latter authors suggest grains as the possible origin of the unexplained portion of CN; cf. also \citet{haenni2020}. Alternatively, many volatile candidate molecules have been suggested. However, the origin of CN in comets where HCN cannot be its sole parent remains to be found. Cyanogen, already mentioned by \citet{swings1956}, ranks high among the candidates as it is a simple molecule that photodissociates into two CN radicals. For this reason, and because of the lack of experimental values, the quantum yield of cyanogen photodissociation into CN is often assumed to be equal to two \citep[e.g.,][]{bockelee-morvan1985}. The reported photodissociation rates vary. \citet{bockelee-morvan1985} determined a photodissociation rate of $3.08 \times 10^{-5}$~s$^{-1}$. Other possible candidate molecules that may produce CN upon photodissociation are the two nitrile species CH$_3$CN and HC$_3$N. \citet{fray2005} review the photochemical properties of these two molecules. The CN production rate from CH$_3$CN photodissociation is likely to be 100 times lower than from HCN photodissociation and hence on the order of $1 \times 10^{-7}$~s$^{-1}$. This results from a photodissociation rate of $6.68 \times 10^{-6}$~s$^{-1}$ \citep{bockelee-morvan1985} when the quantum yield is taken into account. For the vacuum ultraviolet region, \citet{kanda1999} reported that CN production is a minor dissociation channel and that the corresponding quantum yield may be lower than 0.02. HC$_3$N seems to have a higher photodissociation rate. Values of between $2.8 \times 10^{-5}$~s$^{-1}$ and $7.7 \times 10^{-5}$~s$^{-1}$ were mentioned by \citet{fray2005}; however, the authors note that the quantum yield may be as low as 0.05 \citep[cf.][]{halpern1988}. Haser model calculations indicate that NCCN/HCN production rate ratios of between 0.15 and 0.85 and HC$_3$N/HCN production rate ratios of between 0.2 and 0.7 (under the assumption of a quantum yield equal to 1) could explain the discrepancies in the observed scale lengths.\footnote{CH$_3$CN and HNC were not considered for Haser modeling by \citet{fray2005} because these molecules do not possess photodissociation rates higher than that of HCN.}\\
When the Rosetta spacecraft reached the inner coma and approached the cometary nucleus of 67P for the first time, a new era of comet studies began. During the two years of the Rosetta science mission phase, from August 2015 through September 2016, the DFMS, a high-resolution sector-field mass spectrometer that was part of the Rosetta Orbiter Spectrometer for Ion and Neutral Analysis (ROSINA) sensor package \citep{balsiger2007}, analyzed the neutral composition of 67P's inner coma from up-close. Comet 67P is a Jupiter-family comet (JFC) with a period of 6.45 years, a perihelion distance of 1.24~au, and an aphelion distance of 5.68~au. The tilted rotation axis of the comet leads to an inbound equinox in May 2015 at 1.7~au and an outbound equinox in March 2016 at 2.7~au. 67P acquired its current orbit after a close encounter with Jupiter in 1959. Although no reliable orbital parameters exist before 1923 (the year of an earlier encounter with Jupiter), 67P is believed to have resided in the inner Solar System for several thousand years \citep{maquet2015}. Rosetta was as close as 10 km from the cometary nucleus for the first time in October 2014. At that time, the northern hemisphere was experiencing a long but cold summer. Illumination conditions gradually changed on the inbound trajectory, giving way to a short and intense summer on the southern hemisphere with a peak in outgassing shortly after 67P's closest approach to the Sun.\\
Bulk abundances of CH$_3$CN and HC$_3$N relative to water were derived by \citet{rubin2019} from data collected in May 2015 (at approximately 1.6 au) over the southern hemisphere: (0.0059 $\pm$ 0.0034)\% and (0.00040 $\pm$ 0.00023)\%, respectively. These species are frequently observed in comets \citep[e.g.,][]{bockelee-morvan2017}. For cometary cyanogen, mentioned for the first time in \citet{altwegg2019}, no abundance was derived for 67P, or for any other comet previously. Despite intense searches, until now, Titan's atmosphere is the only other Solar System environment where this molecule has been identified \citep[e.g.,][]{kunde1981,teanby2006,teanby2009}.\\
In this article, we derive an abundance of cometary cyanogen relative to hydrogen cyanide for the first time, namely from in situ data collected in the inner coma of comet 67P by the ROSINA/DFMS. The strategy pursued regarding data analysis is described in Sect.~\ref{sec:instr}. Subsequently, the results are presented in Sect.~\ref{sec:res} together with the full mission data for cyanoacetylene and acetonitrile and in Sect.~\ref{sec:disc} we discuss these in the context of values published for other astrophysical environments.

\section{Instrumentation and method}\label{sec:instr}
The DFMS and its fully identical laboratory twin were built in the Mattauch-Herzog geometry \citep{mattauch1934}. Details and specifications of the DFMS were given in \citet{balsiger2007}. The instrument was designed to analyze cometary neutrals after electron-impact ionization (EI) with a 45 eV electron beam. Because of the ionization process, the molecules often undergo fragmentation, which has to be taken into account, for instance when neutral density data are to be derived. Although mass spectrometry does in principle allow to distinguish structural isomers based on their fragmentation patterns, which may be different due to the different structural entities of the molecules, this distinction is not possible for nitrile species, primarily because of the lack of reference data. Nitrile species have not been calibrated in the DFMS laboratory twin model at the University of Bern for reasons of toxicity. Consequently, the fragmentation analysis we performed relies on mass spectra from the National Institute of Standards and Technology (NIST) Standard Reference Database Number 69 \citep{NIST} and hence on data acquired with a standard 70 eV ionizing electron beam. It should be noted that lower electron energies (at equal charge density) lead to lower fragment yields. For C$_3$HN, only data for cyanoacetylene are available.\footnote{The Hill notation for chemical formulas is used in this text to denote a molecule with several structural isomers if no specific isomer is addressed, for instance because distinction based on the fragmentation patterns is not possible.} For C$_2$H$_3$N, spectra of two isomers are available, namely acetonitrile (CH$_3$CN) and methylisocyanide (CH$_3$NC). However, these are very similar and can barely be distinguished, especially as the different ionization voltage would expectedly impose (slightly) different signal intensities. The analysis presented here is therefore based on the NIST fragmentation patterns of acetonitrile and cyanoacetylene, which also are the most common isomers for comets \citep[cf. e.g.][]{mumma2011}. Also, for the full mission data for CHN, the fragmentation of the more stable isomer hydrogen cyanide (HCN) is used. Notably, no fragmentation data are available for its isomer hydrogen isocyanide (HNC). For C$_2$N$_2$, only the mass spectrum of NCCN was measured by \citet{stevenson1950}. NCCN produces a main (molecular ion) peak on \textit{m/z} = 52 u/e and only small signals on other masses, which are due to fragments and/or isotopologs (roughly 10\% combined). Among the four possible structural isomers of C$_2$N$_2$, the most stable is NCCN. \citet{botschwina1990} reported CNCN and CNNC to be less stable than NCCN by 102 and 302 kJ$\cdot$mol$^{-1}$, respectively. We therefore consider it reasonable to attribute the C$_2$N$_2$ signal to NCCN. However, we chose not to correct for fragmentation effects and directly compare the NCCN signal on \textit{m/z} = 52 u/e (corresponding to 90\% of the total intensity due to NCCN) to the HCN signal on \textit{m/z} = 27 u/e (corresponding to 84\% of the total intensity due to HCN).

   \begin{figure}
   \centering
   \includegraphics[width=8cm]{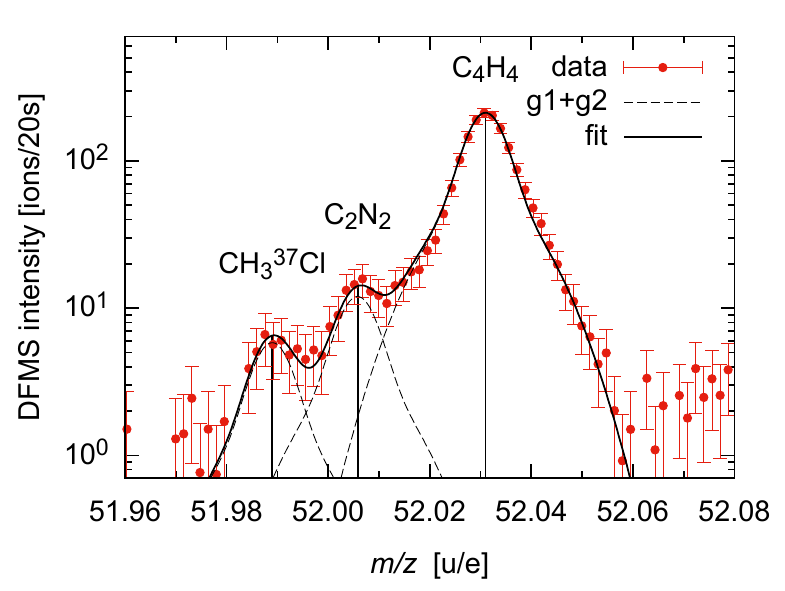}
      \caption{Example mass spectrum collected on 15 March 2016 (10:59 am) of mass 52 u/e --the mass of the molecular ion of NCCN-- in the DFMS neutral gas mode. The sum-curve of the two Gaussians (g1+g2) is plotted with a dashed black line for each peak while the overall fit is shown with a solid black line. Error bars indicate statistical error margins.}
         \label{fig:spec}
   \end{figure}

\noindent The data evaluation regarding NCCN had to be performed manually spectrum by spectrum, based on a least-squares fitting routine. This is because the signal intensity was low, often even below detection limit. Moreover, the peak position is overlapping with the CH$_3$$^{37}$Cl signal to the left and with the C$_3$H$_2$N and also the relatively large C$_4$H$_4$ signals on the right. Each spectrum has been background-subtracted and mass-scaled prior to fitting. The characteristic peak shape of DFMS is best reproduced with a double-Gaussian peak profile when the second Gaussian (g2) is about 5-10\% of the height of the main Gaussian (g1) and about 300-400\% of its width \citep{leroy2015,dekeyser2019}. Width and height ratios of the two Gaussians remain constant within one mass spectrum. A typical mass spectrum with an NCCN signal on mass 52 u/e is shown in Fig.~\ref{fig:spec}. For reasons discussed in Sect.~\ref{sec:res}, the abundance of NCCN was deduced relative to HCN as a neutral density ratio $n_\mathrm{NCCN}/n_\mathrm{HCN}$. This ratio can be obtained from the ratio of the observed ions per spectrum $c_\mathrm{NCCN}/c_\mathrm{HCN}$ as follows:

\begin{equation}
\label{eqn:corrections}
\frac{c_\mathrm{NCCN}}{c_\mathrm{HCN}} \cdot \left(\frac{m_\mathrm{HCN}}{m_\mathrm{NCCN}}\right)^{-0.82} \cdot \frac{\sigma_\mathrm{HCN}}{\sigma_\mathrm{NCCN}} = \frac{n_\mathrm{NCCN}}{n_\mathrm{HCN}},
\end{equation}

\noindent where $\sigma$ denotes the electron-impact ionization cross-section and $m$ denotes the mass of the respective species. The factor $(m_\mathrm{HCN}/m_\mathrm{NCCN})^{-0.82}$ is an empirical correction ---established based on noble gas calibration measurements with the DFMS laboratory twin instrument---that is meant  to account for the mass-dependent instrument sensitivity and detector yield \citep{calmonte2015,wurz2015,rubin2019}. The error related to this correction is estimated to be 15-20\%. Because of the toxicity of the substances under consideration, EI cross-sections were taken from \citet{pandya2012} for HCN (approximately 3.4 \AA$^2$ at 45 eV) and NCCN (5.1 \AA$^2$ at 45 eV). These values yield $\sigma_\mathrm{HCN}/\sigma_\mathrm{NCCN} = 0.66$, a term used in Eq.~\ref{eqn:corrections}. This ratio is expected to have low errors as the cross-sections were calculated based on the same method. The respective Binary Encounter Bethe (BEB) values are 4.5 \AA$^2$ for NCCN and 3.0 \AA$^2$ for HCN, also with a ratio of $\sigma_\mathrm{HCN}/\sigma_\mathrm{NCCN} = 0.66$.\\
The total statistical error of the neutral density ratio $n_\mathrm{NCCN}/n_\mathrm{HCN}$ is governed by the relative statistical error of the NCCN signal, while the HCN signal, being orders of magnitude larger, has a negligible contribution.\footnote{All statistical errors are given as $1\sigma$ errors.} For very small signals, a 5\% error must be added to the statistical error to account for background-subtraction errors and another 10\% error to account for fitting errors. Uncertainties related to fragmentation effects and ionization cross-sections are systematic and the according error is estimated to not exceed 20\% in total. HCN is normally measured at a lower gain step than the minor species, which adds a 10\% systematic error to the ratios \citep{schroeder2019}. Systematic errors are taken into account in Table~\ref{tab:comp}, where the total error margins are given, but are neglected elsewhere.\\
The data evaluation procedure applied to derive HC$_3$N and CH$_3$CN neutral densities was described in great detail by \citet{rubin2019} and in the literature cited therein. Unlike for NCCN, the number of ions per spectrum, which was used as a starting point for the analysis presented in this work, was derived without applying a mass scale. Fragmentation effects have also been considered. The statistical error of the neutral density of a moderately abundant species like HCN is approximately 10\%. In order to obtain the total error, the systematic errors indicated above have to be included.

\section{Results}\label{sec:res}

   \begin{figure*}[h!t]
   \centering
   \includegraphics[width=\hsize]{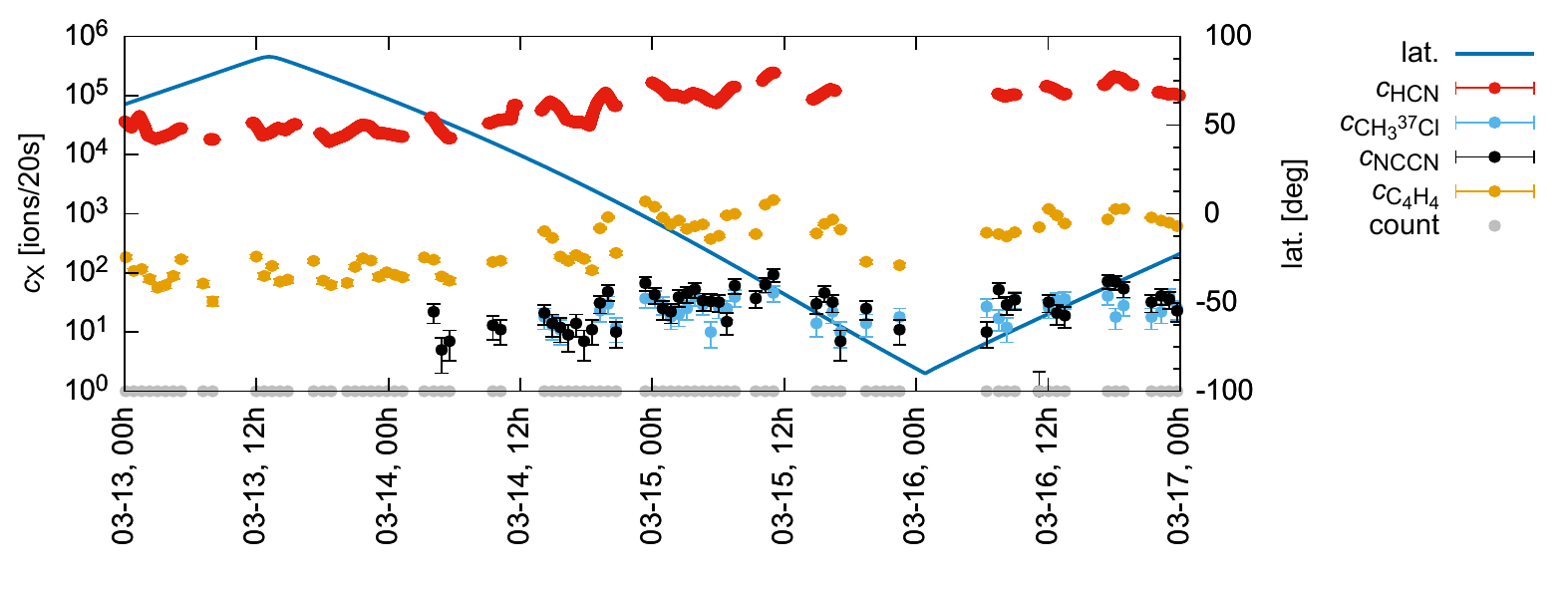}
   \caption{Number of ions per spectrum $c_\mathrm{X}$, where X denotes the species observed on mass 52 u/e (i.e., CH$_3$$^{37}$Cl, NCCN, and C$_4$H$_4$). The underlying data were collected in the four days between 13 and 16 March 2016, including the day of the outbound equinox on 16 March 2016. Counting symbols on the x-axis mark times of analyzed spectra. HCN data has been evaluated in an automated data evaluation pipeline and time-interpolated as described in \citet{rubin2019}. Error bars indicate statistical errors.}
                 \label{fig:mar16}
    \end{figure*}

\begin{figure*}[h!t]
   \centering
   \includegraphics[width=\hsize]{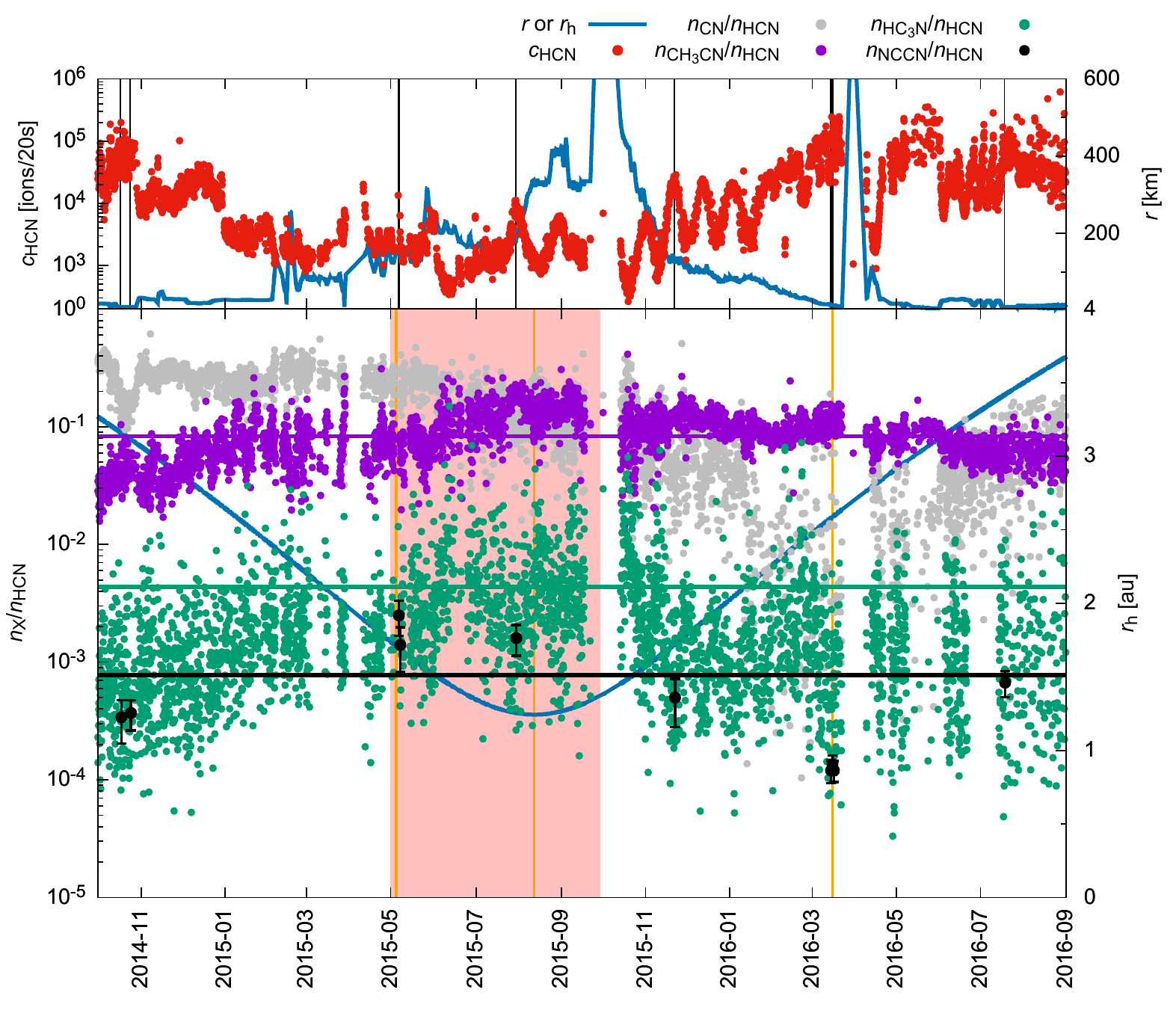}
   \caption{(top) Number of ions per spectrum of HCN ($c_\mathrm{HCN}$; with 30\% error bars and fragmentation not taken into account) during the full Rosetta science mission as derived by \citet{rubin2019}. The cometocentric distance ($r$) is over-plotted (solid blue line) together with the days when NCCN was found (black vertical bars). (bottom) Neutral density ratio of species X with respect to HCN ($n_\mathrm{X}/n_\mathrm{HCN}$), where X is CH$_3$CN, HC$_3$N, NCCN, and CN (from \citet{haenni2020} for comparison). The respective total mission average values are included as solid horizontal lines for the nitriles investigated in this work. The inbound equinox (5 May 2015), perihelion (13 August 2015), and the outbound equinox (16 March 2016) are indicated (orange vertical bars) together with the heliocentric distance ($r_\mathrm{h}$; solid blue line). The period from May to September 2015 used to derive the close-to-perihelion values in Table~\ref{tab:comp} is indicated in pale red. The data shown here for cyanogen are also listed in Table~\ref{tab:app} in Appendix~\ref{app:2}. Statistical error bars are omitted for visual clarity (except for NCCN). An impression can be obtained for the case of $n_\mathrm{CN}/n_\mathrm{HCN}$ from Fig. 4 (top) in \citet{haenni2020}.}
                 \label{fig:fullmiss-hcn}
    \end{figure*}

In several spectra collected between 13 and 16 March 2016, just when 67P reached its outbound equinox, the mass spectrometric signature of NCCN was clearly observed. Notably,  the abundance of species may be highly variable on short timescales, for example due to the spacecraft trajectory or due to variability of the coma itself. This is obvious from Fig.~\ref{fig:mar16}, which shows the uncorrected number of ions collected on mass 52 u/e during the indicated time period. On 13 March, the signature of NCCN is not detectable and also the C$_4$H$_4$ signal is often rather weak. The situation completely changes within a day and the signature of NCCN becomes clearly apparent. Based on these data we investigated correlations with other coma species. The resulting Pearson correlation coefficients are listed in Table~\ref{tab:correl} in the Appendix~\ref{app:1}. Relatively good correlation (R = 0.8 or higher) of the NCCN signature is observed with only a few species, HCN among them. For most of the other species, for which we have data available, the correlation is either weak or nonexistent. We therefore chose to use HCN as a reference species. There may be various reasons why  the NCCN and the HCN signals correlate quite well along longitude and latitude of the comet, such as the high volatility, the shared nitrile functionality, or maybe even a similar formation route from CN. \citet{agundez2018} recapitulate that both CNCN (not barrierless) and NCCN (barrierless) in dark clouds may be formed from CN + HNC (the reaction CN + HCN is known to possess an energy barrier). As observed for other comets \citep[e.g.,][]{mumma2011}, it is likely that also for 67P a (presumably small) portion of the HCN observed by DFMS is actually HNC. For reasons detailed in Sect.~\ref{sec:instr}, structural isomers with identical exact masses cannot always be distinguished and we therefore assumed the observed molecules to have the structure of the most stable isomer.\\
\noindent Thanks to the good correlation of NCCN and HCN, we were able to successfully locate periods where the NCCN signature was detectable, namely by looking at periods where the HCN abundance peaks. This is illustrated in Fig.~\ref{fig:fullmiss-hcn} (top), which shows the HCN signal $c_\mathrm{HCN}$ during the full Rosetta mission time. Days where the signature of NCCN was found are indicated by black bars. Data from these days were evaluated as described in Sect.~\ref{sec:instr} in order to extract the neutral density ratio $n_\mathrm{NCCN}/n_\mathrm{HCN}$ as shown in Fig.~\ref{fig:fullmiss-hcn} (bottom). The full mission data of the neutral abundance ratios $n_\mathrm{CH_3CN}/n_\mathrm{HCN}$ and $n_\mathrm{HC_3N}/n_\mathrm{HCN}$ are included. Although the correlation of NCCN and HCN was good in March 2016, the dependence on heliocentric distance is obviously different for the two species, which is a phenomenon widely known for various cometary species. However, based on mass spectrometric data we cannot fully exclude that, first, the signal assigned to HCN is only due to hydrogen cyanide and does not contain contributions from hydrogen isocyanide and, second, a possible contribution of the isocyanide would be time-independent. A time-variable contribution of hydrogen isocyanide (if it fragments differently from its isomer) in principle could be responsible for such variations of the abundances relative to HCN to a certain extent \citep[cf. e.g.,][]{mumma2011}. The same is true for possible contributions of isomers of the other nitrile species. Also, the CH$_3$CN relative abundance, which generally displays less variation because of a better correlation with HCN, clearly varies around the average value of the $n_\mathrm{CH_3CN}/n_\mathrm{HCN}$ ratio of 0.083 with peak values around 67P's perihelion passage. The same trend is observed for both of the other nitrile species, HC$_3$N and NCCN, which vary more pronouncedly. While ratios around perihelion passage are clearly above average (average $n_\mathrm{HC_3N}/n_\mathrm{HCN} = 0.0044$; average $n_\mathrm{NCCN}/n_\mathrm{HCN} = 0.00078$), ratios far away from it are rather below average. The difference between the high and low values is more than an order of magnitude. The large scatter of the $n_\mathrm{HC_3N}/n_\mathrm{HCN}$ values for 67P, indicating the comparably poor correlation with HCN, leads to identical averages for the full mission as compared to the time around perihelion. The average relative abundances of the nitrile species presented in this work are summarized in Table~\ref{tab:comp}, including bulk abundances from \citet{rubin2019} where available, as well as values reported for other astrophysical environments (to be discussed in Sect.~\ref{sec:disc}). The bulk values from \citet{rubin2019} represent the outgassing as observed between the end of May and beginning of June 2015. The values for $n_\mathrm{HC_3N}/n_\mathrm{HCN}$ found by \citet{rubin2019} are consistent within error limits with the averages presented in this work, while their $n_\mathrm{CH_3CN}/n_\mathrm{HCN}$ 
is clearly falling below. Still far from the Sun (at approximately 3.1 au in November 2014), \citet{leroy2015} derived a ratio of CH$_3$CN/HCN of 0.066 for the northern hemisphere and 0.025 for the southern hemisphere, which bracket the bulk value from \citet{rubin2019}. At that time, only an upper limit for HC$_3$N could be derived. However, momentary outgassing, as captured for instance by \citet{rubin2019} or \citet{leroy2015}, may well differ from a full mission average and also from the integral over the full mission time, as presented by \citet{laeuter2020}. The latter authors derived abundances of 14 major coma species relative to water based on an inverse coma model and ROSINA/DFMS data. Their Fig. 5 compares their findings to the relative abundances from \citet{rubin2019} (amongst others), demonstrating reasonable consistency.

\begin{table*}[h!t]
\caption[]{Abundances of the targeted nitrile species relative to HCN as resulting from our analysis of ROSINA/DFMS data from 67P$^{(a)}$ and other astrophysical environments.}
\centering
\begin{tabular}{llll}
\hline\hline
Environment                     & NCCN/HCN                  & HC$_3$N/HCN               & CH$_3$CN/HCN                 \\
\hline
67P (full mission)              & 0.00078 $\pm$ 0.00030     & 0.0044 $\pm$ 0.0009       & 0.083 $\pm$ 0.013            \\
67P (1.24--1.74 au)             & 0.0018 $\pm$ 0.0009       & 0.0046 $\pm$ 0.0008       & 0.12 $\pm$ 0.02              \\
67P (1.5--1.6 au)$^{(b)}$       & --                        & 0.0029 $\pm$ 0.0019       & 0.042 $\pm$ 0.027            \\
\hline
Hale-Bopp (0.9-1.2 au)$^{(c)}$  & --                        & 0.079                     & 0.082                        \\
C/2014 Q2$^{(d)}$               & --                        & 0.022 $\pm$ 0.006         & 0.17 $\pm$ 0.05              \\
\hline                          
Disk average$^{(e)}$            & --                        & 0.057 $\pm$ 0.023         & 0.055 $\pm$ 0.022            \\
IRAS 16293-2422 B$^{(f)}$       & --                        & 0.0036 $\pm$ 0.001        & 0.8 $\pm$ 0.2                \\
   \hline
                \end{tabular}
        \label{tab:comp}
        \tablefoot{$^{(a)}$ Total error margins include statistical errors for the averages calculated as standard deviation divided by the square root of the number of actual measurements in the averaged time period (excluding time-interpolated data points). $^{(b)}$ Values from \citet{rubin2019}. $^{(c)}$ Values from \citet{bockelee-morvan2000}. $^{(d)}$ Values from \citet{biver2015}. $^{(e)}$ Values from \citet{bergner2018}; 40\% error margins estimated from the error bars in their Fig. 12. The average over the sample of five disks was calculated considering rotational temperatures of 30 K (CH$_3$CN/HCN) and 70 K (HC$_3$N/HCN), respectively, not taking into account CH$_3$CN relative abundance upper limits given for two of the five disks. $^{(f)}$ Values from \citet{drozdovskaya2019}.}
\end{table*}

\noindent Based on the data presented in Fig.~\ref{fig:fullmiss-hcn} (bottom), the heliocentric distance dependence of the species targeted in this work can be obtained. This is usually achieved by fitting a power law to the slope of the production rate of a given species as a function of the heliocentric distance $r_h$. The abundance of a species X relative to HCN, as derived in this work, is proportional to its production rate relative to HCN. Therefore, the slope of the production rate of species X relative to HCN corresponds to the difference between the two individual slopes. Using the same time periods $I_a$, $I_b$, and $I_c$ as \citet{laeuter2020}, we obtained the exponents $x_a$, $x_b$, and $x_c$ from fitting the following power law to our data:

\begin{equation}
\label{eqn:power-law}
\frac{n_\mathrm{X}}{n_\mathrm{HCN}} \propto r_h^{x}.
\end{equation}

\begin{table*}[h!t]
\caption[]{Power-law exponents for CH$_3$CN, HC$_3$N, and CN for three time intervals $I$$^{(a)}$.}
\centering
\begin{tabular}{llll}
\hline\hline
                  &    $I_a$                    &    $I_b$                 &    $I_c$                  \\
Species           &    3.1--2.3 au              &    1.7--2.2 au           &    2.4--3.6 au            \\
                  &    11/2014--02/2015         &    11/2015--01/2016      &    02/2016--08/2016       \\
\hline
CH$_3$CN          &   +0.3                      &    -3.6                  &    -3.7                   \\
HC$_3$N           &   +1.2                      &    -9.5                  &    -4.1                   \\
CN                &   +2.3                      &    -6.6                  &    +3.2                   \\
\hline
\end{tabular}
\tablefoot{$^{(a)}$ Further information on the employed time intervals $I_a$, $I_b$, and $I_c$ can be found in Table 3 of \citet{laeuter2020}.}
\label{tab:rh-dep}
\end{table*}

\noindent Adding to $x$ the appropriate power-law exponent for HCN reported in Table 4 of \citet{laeuter2020} yields the desired power-law exponent of the heliocentric distance dependence of our species X, that is of CH$_3$CN and HC$_3$N (albeit with less confidence). The same procedure was applied to the CN data from \citet{haenni2020}. The NCCN data  are unfortunately too limited for such a fitting. However, as $n_\mathrm{NCCN}/n_\mathrm{HCN}$ seems to follow $n_\mathrm{HC_3N}/n_\mathrm{HCN}$ quite well, a similar heliocentric distance dependence can be expected. Table~\ref{tab:rh-dep} summarizes the accordingly derived power-law exponents for the distinct time periods $I$, including those for HCN, H$_2$O, and CO$_2$ found by \citet{laeuter2020}. Notably, the correlation coefficients for period $I_c$ are rather low (R $\approx$ 0.6 for CH$_3$CN and CN, less for HC$_3$N). Nevertheless, they are even lower for the other periods, especially for period $I_a$, which  reflects the large scatter of the data. \citet{laeuter2020} used period $I_c$ to group molecules into two groups, one that follows CO$_2$ and one that rather follows H$_2$O. Using the same criteria, CH$_3$CN and HC$_3$N fall into the H$_2$O group rather than into the CO$_2$ group. This seems to suggest that they are embedded in a water-ice matrix. CN, in contrast, follows a positive exponent in time period $I_c$, making an origin of CN from one of the considered nitrile molecules highly unlikely. However, the positive exponent of CN could be the result of the interdependence between cometocentric and heliocentric distance. For a direct comparison, Fig.~\ref{fig:fullmiss-hcn} also includes the lower limit\footnote{\citet{haenni2020} do not correct for the portion CN lost due to fragmentation inside the instrument because this portion has not been determined for the DFMS.} neutral density ratio $n_\mathrm{CN}/n_\mathrm{HCN}$, as derived and reported by \citet{haenni2020} and used for fitting the power-law exponents listed in Table~\ref{tab:rh-dep}.\\
Returning to the initial question regarding the search for alternative or additional parent species of the CN radical, the following can be stated on the basis of our analysis of relative abundances: in the first few months of the Rosetta science mission, HC$_3$N and NCCN abundances on the order of 0.1\% or less relative to HCN fall far below the minimum relative abundance of 10-20\%, as required by the Haser modeling of \citet{fray2005}. Remember that CH$_3$CN can be ruled out as parent species due to unfavorable photochemical properties, despite its higher abundance of up to about 10\% relative to HCN. Also, abundances of HC$_3$N and NCCN relative to HCN are at least two orders of magnitude lower than the abundance of CN relative to HCN. Judging from the photorates of the targeted species, the assumed gas velocity of a little less than 1 km/s, and the fact that ROSINA/DFMS data were collected at a few tens of kilometers above the cometary surface, parent species cannot be less abundant than daughter species. An exemplary estimation of HCN photodissociation at cometocentric distances relevant for Rosetta and at heliocentric distances relevant for 67P may be found in \citet{haenni2020}. Also, the approximate factor of two between some radio and IR HCN production rates ---compare for example \citet{fray2005} to \citet{dellorusso2016} --- does not substantially change the picture. On the contrary, the fact that the relative CN abundance from \citet{haenni2020} is considered a lower limit rather increases the discrepancy. The fact that the situation gradually changes towards the outbound equinox, where $n_\mathrm{CN}/n_\mathrm{HCN}$ reaches a minimum, is due to the different heliocentric distance dependence of these species relative to HCN (cf. discussion above and Table~\ref{tab:rh-dep}). While the relative abundance of CN decreases, reaching a minimum around the time of the outbound equinox in March 2016, the relative abundances of the nitriles studied in this work peak around perihelion.

\section{Discussion}\label{sec:disc}
Having derived a relative abundance of cyanogen for the first time and having followed this and the relative abundances of two other nitrile species during the full Rosetta mission phase, this section compares our results to other astrophysical environments where the targeted species were observed. A selection of values can be found in Table~\ref{tab:comp}. For the following discussion, it is important to remember that absolute abundances of molecules with multiple isomers derived from ROSINA/DFMS data should be considered as upper limits. This is due to the fact that other isomers, besides the most likely one considered in the data evaluation processes for this work, may contribute to the observed signal. However, because this is the case for all of the investigated nitrile species (incl. HCN), the effect is probably compensated at least partially when abundance ratios are considered.

\subsection{Comparison to other comets}\label{subsec:comets}
Unlike the smaller and less active JFC 67P, the large and active Oort cloud comet (OCC) Hale-Bopp shows very similar relative abundances of HC$_3$N and CH$_3$CN. HC$_3$N is less abundant in 67P than in Hale-Bopp, while CH$_3$CN is more abundant close to the Sun. Nevertheless, the  full mission average of this latter species matches  the value from Hale-Bopp exactly. Hale-Bopp was observed between 0.9 and 1.2 au and hence the solar flux must have been considerably higher than for 67P between 1.24 and 1.74 au. This, together with the heliocentric distance dependence of $n_\mathrm{HC_3N}/n_\mathrm{HCN}$, could (at least partially) explain the higher HC$_3$N/HCN ratio for Hale-Bopp (at about 1 au) than for 67P (at about 1.5 au). HCN itself is less abundant relative to water in 67P than in Hale-Bopp by almost a factor of two, comparing \citet{rubin2019} and \citet{bockelee-morvan2000}. Notably, the HCN abundance measured with radio spectroscopy is lower than the one measured with IR spectroscopy by a factor 1.4 for Hale-Bopp, meaning that CH$_3$CN/HCN and HC$_3$N/HCN may be overestimated (assuming that other nitriles are not similarly biased).\\
Another comet for which the relevant abundance ratios have been determined is the OCC C/2014 Q2 (Lovejoy). Using the 30-m telescope of the Institut de Radioastronomie Millimétrique (IRAM), \citet{biver2015} observed this comet at a heliocentric distance of 1.3 au, very similar to the perihelion distance of 67P. The authors identified 21 molecules and retrieved production rates\footnote{The model used by \citet{biver2015} to retrieve production rates includes neutral and electron collisions.}, including those of HCN, CH$_3$CN, and HC$_3$N. The CH$_3$CN/HCN ratio in comet C/2014 Q2 in particular closely matches that in 67P. However, the HC$_3$N/HCN ratio found by these latter authors is roughly double that found in this work.\\
\citet{bockelee-morvan2017} review abundances of the targeted nitriles relative to water by giving a range for a sample of more than ten comets. Interestingly, the range of relative abundances reported for HC$_3$N/H$_2$O includes and exceeds the range of relative abundances reported for CH$_3$CN/H$_2$O. This overall picture does not change when the comparably narrow range of values of HCN abundances relative to water is taken into account: There is a (smaller) portion of comets where HC$_3$N is more abundant than CH$_3$CN and a (larger) portion where it is similarly or less abundant. 67P and C/2014 Q2 clearly belong to the latter group of comets, while Hale-Bopp is borderline. The cause of the large scatter of the relative abundance values of HC$_3$N in comets is under debate.

\subsection{Comparison to the interstellar medium}\label{subsec:disks}
A similar scenario was deduced from observations of a sample of protoplanetary disks by \citet{bergner2018}: while the abundances of CH$_3$CN relative to HCN cover a narrow range of roughly 2 to 6 per cent, those of HC$_3$N relative to HCN cover a broad range of a few to more than 120 per cent (cf. their Fig. 12). For CH$_3$CN/HCN, the assumption of lower rotational temperatures leads to values more similar to what is observed for comets, while for HC$_3$N/HCN the opposite is the case. This leads to disk average abundances of CH$_3$CN relative to HCN and for a rotational temperature of 30 K that are comparable to the Rosetta full mission average reported here. The average from the comet's southern hemisphere (that is from around the perihelion of  67P) is somewhat higher. However, for HC$_3$N/HCN,  even the values for the highest rotational temperature of 70 K are still an order of magnitude higher than the values we report for 67P. This depletion of HC$_3$N relative to HCN, when comparing 67P to Bergner's disk sample, remains unexplained, although it is observed for other comets, such as for instance for comet C/2014 Q2. The relative abundances of the two nitriles in the comet Hale-Bopp are close to 8\%, thus rather compatible with Bergner's disk sample, at least when opposing rotational temperatures are assumed as is the case for the disk averages shown in Table~\ref{tab:comp}. Notably, \citet{bergner2018} mention that the rotational temperature effectively measured for HC$_3$N/HCN in one of the disks is 50 K, making it reasonable that a rotational temperature of 30 K leads to overestimation of the ratio. However, comparison of the nitrile abundances in  67P to those in the vicinity of the low-mass protostar IRAS 16293-2422 B \citep{drozdovskaya2019}  leads us to the conclusion that, while HC$_3$N/HCN is similarly abundant, CH$_3$CN/HCN is overabundant in the protostar. Generally, comparison to disk observations is limited by the fact that often the midplane, where comets are most likely formed, cannot be observed due to optical thickness.\\
Unfortunately, direct remote spectroscopic observations of NCCN are hampered by its lack of allowed rotational transitions and very low vibrational band strength \citep{crovisier1987}. Therefore, \citet{agundez2015,agundez2018} proposed the use of NCCNH$^+$ (protonated cyanogen) or CNCN (isocyanogen) as proxies. Based on the detection of CNCN, NCCNH$^+$, C$_3$N, CH$_3$CN, C$_2$H$_3$CN, and H$_2$CN in the prestellar core L1544, \citet{vastel2019} developed a chemical network to explain the observed abundances. The main cyanogen production pathways considered in the network are the CN + HNC and N + C$_3$N reactions in the gas phase. Their calculations yield an NCCN gas phase abundance a factor 100 larger than that of its isomer CNCN. Although the modeled column density of NCCN in the gas phase is comparable to that of HCN, its modeled grain surface abundance seems to be relatively low. Judging from what we learned about comet 67P, a relatively low grain surface abundance of NCCN with respect to HCN seems reasonable.

\subsection{Comparison to Titan's atmosphere}\label{subsec:titan}
The detection of cyanogen in Titan's atmosphere by the IR spectrometer IRIS onboard Voyager 1 \citep{kunde1981} was later confirmed based on data from the Cassini mission \citep[e.g.,][]{teanby2006,teanby2009,magee2009,cui2009}. Abundances of cyanogen, and of HCN, HC$_3$N, and CH$_3$CN, which are also observable with Atacama Large Millimeter/submillimeter Array \citep[ALMA; e.g.,][]{thelen2019,cordiner2019,lellouch2019}, were consistently observed to underlay strong latitudinal and vertical variations. In Titan's atmosphere, a specific nitrogen(N$_2$)- and methane(CH$_4$)-dominated chemistry is at play. Ultraviolet photons from the Sun, photoelectrons, and magnetospheric electrons dissociate molecular nitrogen and methane in the mesosphere and thermosphere into radicals, which may recombine to form nitriles \citep{lara1996,wilson2004}. \citet{loison2015} proposed the formation of NCCN from HCCN + N, in addition to the previous reactions. A photochemical network of the neutral reactions yielding nitrile and imine species is shown in their Fig. 8. Thanks to ROSINA/DFMS, onboard the Rosetta spacecraft, comet 67P has now become the second environment in our Solar System where cyanogen has been firmly detected. Obviously, we cannot expect to observe neutral--neutral or neutral--ion coma chemistry at 67P. This is because (i) the densities at cometocentric distances of a few tens of kilometers are too low (and hence the mean free path too large) for a relevant amount of collisions to occur, and (ii) the sublimating species have little time to be photoionized on their outbound trajectory up to such distances. But also at the surface, chemistry seems unlikely \citep{fuselier2015,fuselier2016}.


\section{Summary and Conclusions}\label{sec:sumconc}
The most relevant findings of this work can be summarized as follows:\\

   \begin{enumerate}
            \item A time-variable signal of C$_2$N$_2$ was identified in ROSINA/DFMS data. Assuming the most stable isomer to be responsible for this signal, it may be assigned to cometary cyanogen NCCN.
            \item NCCN correlates well with CHN (presumably HCN) in March 2016. Thanks to this correlation, several other periods with NCCN signals above the detection limit were identified and abundances of NCCN relative to HCN were derived. Within 1.24-1.74 au, the average ratio $n_\mathrm{NCCN}/n_\mathrm{HCN}$ is equal to 0.0018 $\pm$ 0.0009.
                        \item Relative abundances of two other nitriles, namely HC$_3$N and CH$_3$CN, were derived with respect to HCN as detailed in \citet{rubin2019}. The average ratio $n_\mathrm{HC_3N}/n_\mathrm{HCN}$ is equal to 0.0046 $\pm$ 0.00046 within 1.24-1.74 au, while the average ratio $n_\mathrm{CH_3CN}/n_\mathrm{HCN}$ is equal to 0.12 $\pm$ 0.012.
                        \item All of the abundances (relative to HCN) are higher around 67P's perihelion passage and lower farther away from it. The various species show different heliocentric distance dependencies. Furthermore, effects of possible contributions of other structural isomers (which may fragment differently) cannot be fully ruled out.
      \item Neither NCCN nor HC$_3$N are abundant enough (relative to HCN) to explain the CN radical density (relative to HCN) observed with ROSINA/DFMS and reported by \citet{haenni2020}. The search for CN parent species has to be continued.
   \end{enumerate}

Presenting evidence against several nitriles as possible parents of the cyano radical in comet 67P, this work may help future investigations of the origin of the cometary CN to pursue more fruitful leads.


\begin{acknowledgements}
We gratefully acknowledge the work of the many engineers, technicians and scientists involved in the Rosetta mission and in the ROSINA instrument in particular. Without their contributions, ROSINA would not have produced such outstanding results. Rosetta is an ESA mission with contributions from its member states and NASA. Work at the University of Bern was funded by the Canton of Bern and the Swiss National Science Foundation (200020 182418). S.F.W. acknowledges the financial support of the SNSF Eccellenza Professorial Fellowship PCEFP2\_181150. Especially, we thank Dr. N. Fray and Dr. M. N. Drozdovskaya for engaging in fruitful discussions on the topic.
\end{acknowledgements}

\bibliographystyle{aa} 
\bibliography{2020_Haenni_C2N2} 

\begin{thebibliography}{53}
\expandafter\ifx\csname natexlab\endcsname\relax\def\natexlab#1{#1}\fi

\bibitem[{{Ag{\'u}ndez} {et~al.}(2015){Ag{\'u}ndez}, {Cernicharo}, {de
  Vicente}, {Marcelino}, {Roueff}, {Fuente}, {Gerin}, {Gu{\'e}lin}, {Albo},
  {Barcia}, {Barbas}, {Bola{\~n}o}, {Colomer}, {Diez}, {Gallego},
  {G{\'o}mez-Gonz{\'a}lez}, {L{\'o}pez-Fern{\'a}ndez},
  {L{\'o}pez-Fern{\'a}ndez}, {L{\'o}pez-P{\'e}rez}, {Malo}, {Serna}, \&
  {Tercero}}]{agundez2015}
{Ag{\'u}ndez}, M., {Cernicharo}, J., {de Vicente}, P., {et~al.} 2015, \aap,
  579, L10

\bibitem[{{Ag{\'u}ndez} {et~al.}(2018){Ag{\'u}ndez}, {Marcelino}, \&
  {Cernicharo}}]{agundez2018}
{Ag{\'u}ndez}, M., {Marcelino}, N., \& {Cernicharo}, J. 2018, \apjl, 861, L22

\bibitem[{{A'Hearn} {et~al.}(1983){A'Hearn}, {Millis}, {Festou}, {Benvenuti},
  {Cacciari}, {Cassatella}, {Talavera}, {Wamsteker}, {Green}, {Hale}, \&
  {Marsden}}]{ahearn1983}
{A'Hearn}, M., {Millis}, R., {Festou}, M., {et~al.} 1983, \iaucirc, 3802, 1

\bibitem[{{Altwegg} {et~al.}(2019){Altwegg}, {Balsiger}, \&
  {Fuselier}}]{altwegg2019}
{Altwegg}, K., {Balsiger}, H., \& {Fuselier}, S.~A. 2019, \araa, 57, 113

\bibitem[{{Balsiger} {et~al.}(2007){Balsiger}, {Altwegg}, {Bochsler},
  {Eberhardt}, {Fischer}, {Graf}, {J{\"a}ckel}, {Kopp}, {Langer}, {Mildner},
  {M{\"u}ller}, {Riesen}, {Rubin}, {Scherer}, {Wurz}, {W{\"u}thrich}, {Arijs},
  {Delanoye}, {de Keyser}, {Neefs}, {Nevejans}, {R{\`e}me}, {Aoustin},
  {Mazelle}, {M{\'e}dale}, {Sauvaud}, {Berthelier}, {Bertaux}, {Duvet},
  {Illiano}, {Fuselier}, {Ghielmetti}, {Magoncelli}, {Shelley}, {Korth},
  {Heerlein}, {Lauche}, {Livi}, {Loose}, {Mall}, {Wilken}, {Gliem}, {Fiethe},
  {Gombosi}, {Block}, {Carignan}, {Fisk}, {Waite}, {Young}, \&
  {Wollnik}}]{balsiger2007}
{Balsiger}, H., {Altwegg}, K., {Bochsler}, P., {et~al.} 2007, \ssr, 128, 745

\bibitem[{{Bergner} {et~al.}(2018){Bergner}, {Guzm{\'a}n}, {{\"O}berg},
  {Loomis}, \& {Pegues}}]{bergner2018}
{Bergner}, J.~B., {Guzm{\'a}n}, V.~G., {{\"O}berg}, K.~I., {Loomis}, R.~A., \&
  {Pegues}, J. 2018, \apj, 857, 69

\bibitem[{{Biver} {et~al.}(2015){Biver}, {Bockel{\'e}e-Morvan}, {Moreno},
  {Crovisier}, {Colom}, {Lis}, {Sandqvist}, {Boissier}, {Despois}, \&
  {Milam}}]{biver2015}
{Biver}, N., {Bockel{\'e}e-Morvan}, D., {Moreno}, R., {et~al.} 2015, Science
  Advances, 1, 1500863

\bibitem[{{Bockelée-Morvan} {et~al.}(1984){Bockelée-Morvan}, {Crovisier},
  {Baudry}, {Despois}, {Perault}, {Irvine}, {Schloerb}, \&
  {Swade}}]{bockelee-morvan1984}
{Bockelée-Morvan}, D., {Crovisier}, J., {Baudry}, A., {et~al.} 1984, \aap,
  141, 411

\bibitem[{{Bockel{\'e}e-Morvan} \& {Biver}(2017)}]{bockelee-morvan2017}
{Bockel{\'e}e-Morvan}, D. \& {Biver}, N. 2017, Philosophical Transactions of
  the Royal Society of London Series A, 375, 20160252

\bibitem[{{Bockel{\'e}e-Morvan} \& {Crovisier}(1985)}]{bockelee-morvan1985}
{Bockel{\'e}e-Morvan}, D. \& {Crovisier}, J. 1985, \aap, 151, 90

\bibitem[{{Bockel{\'e}e-Morvan} {et~al.}(2000){Bockel{\'e}e-Morvan}, {Lis},
  {Wink}, {Despois}, {Crovisier}, {Bachiller}, {Benford}, {Biver}, {Colom},
  {Davies}, {G{\'e}rard}, {Germain}, {Houde}, {Mehringer}, {Moreno}, {Paubert},
  {Phillips}, \& {Rauer}}]{bockelee-morvan2000}
{Bockel{\'e}e-Morvan}, D., {Lis}, D.~C., {Wink}, J.~E., {et~al.} 2000, \aap,
  353, 1101

\bibitem[{{Botschwina} \& {Sebald}(1990)}]{botschwina1990}
{Botschwina}, P. \& {Sebald}, P. 1990, Chemical Physics, 141, 311

\bibitem[{{Calmonte}(2015)}]{calmonte2015}
{Calmonte}, U.~M. 2015, {Sulfur isotopic ratios at 67P/Churyumov-Gerasimenko
  and characterization of ROSINA-DFMS FM and FS}

\bibitem[{{Cordiner} {et~al.}(2019){Cordiner}, {Teanby}, {Nixon}, {Vuitton},
  {Thelen}, \& {Charnley}}]{cordiner2019}
{Cordiner}, M.~A., {Teanby}, N.~A., {Nixon}, C.~A., {et~al.} 2019, \aj, 158, 76

\bibitem[{{Crovisier}(1987)}]{crovisier1987}
{Crovisier}, J. 1987, \aaps, 68, 223

\bibitem[{{Crovisier}(1994)}]{crovisier1994}
{Crovisier}, J. 1994, \jgr, 99, 3777

\bibitem[{{Cui} {et~al.}(2009){Cui}, {Yelle}, {Vuitton}, {Waite}, {Kasprzak},
  {Gell}, {Niemann}, {M{\"u}ller-Wodarg}, {Borggren}, {Fletcher}, {Patrick},
  {Raaen}, \& {Magee}}]{cui2009}
{Cui}, J., {Yelle}, R.~V., {Vuitton}, V., {et~al.} 2009, \icarus, 200, 581

\bibitem[{{De Keyser} {et~al.}(2019){De Keyser}, {Altwegg}, {Gibbons},
  {Dhooghe}, {Balsiger}, {Berthelier}, {Fuselier}, {Gombosi}, {Neefs}, \&
  {Rubin}}]{dekeyser2019}
{De Keyser}, J., {Altwegg}, K., {Gibbons}, A., {et~al.} 2019, International
  Journal of Mass Spectrometry, 446, 116232

\bibitem[{{Dello Russo} {et~al.}(2016){Dello Russo}, {Kawakita}, {Vervack}, \&
  {Weaver}}]{dellorusso2016}
{Dello Russo}, N., {Kawakita}, H., {Vervack}, R.~J., \& {Weaver}, H.~A. 2016,
  \icarus, 278, 301

\bibitem[{{Dello Russo} {et~al.}(2009){Dello Russo}, {Vervack}, {Weaver},
  {Kawakita}, {Kobayashi}, {Biver}, {Bockel{\'e}e-Morvan}, \&
  {Crovisier}}]{dellorusso2009}
{Dello Russo}, N., {Vervack}, R.~J., J., {Weaver}, H.~A., {et~al.} 2009, \apj,
  703, 187

\bibitem[{{Drozdovskaya} {et~al.}(2019){Drozdovskaya}, {van Dishoeck}, {Rubin},
  {J{\o}rgensen}, \& {Altwegg}}]{drozdovskaya2019}
{Drozdovskaya}, M.~N., {van Dishoeck}, E.~F., {Rubin}, M., {J{\o}rgensen},
  J.~K., \& {Altwegg}, K. 2019, \mnras, 490, 50

\bibitem[{{Fray} {et~al.}(2005){Fray}, {B{\'e}nilan}, {Cottin}, {Gazeau}, \&
  {Crovisier}}]{fray2005}
{Fray}, N., {B{\'e}nilan}, Y., {Cottin}, H., {Gazeau}, M.~C., \& {Crovisier},
  J. 2005, \planss, 53, 1243

\bibitem[{{Fuselier} {et~al.}(2016){Fuselier}, {Altwegg}, {Balsiger},
  {Berthelier}, {Beth}, {Bieler}, {Briois}, {Broiles}, {Burch}, {Calmonte},
  {Cessateur}, {Combi}, {De Keyser}, {Fiethe}, {Galand}, {Gasc}, {Gombosi},
  {Gunell}, {Hansen}, {H{\"a}ssig}, {Heritier}, {Korth}, {Le Roy},
  {Luspay-Kuti}, {Mall}, {Mandt}, {Petrinec}, {R{\`e}me}, {Rinaldi}, {Rubin},
  {S{\'e}mon}, {Trattner}, {Tzou}, {Vigren}, {Waite}, \& {Wurz}}]{fuselier2016}
{Fuselier}, S.~A., {Altwegg}, K., {Balsiger}, H., {et~al.} 2016, \mnras, 462

\bibitem[{{Fuselier} {et~al.}(2015){Fuselier}, {Altwegg}, {Balsiger},
  {Berthelier}, {Bieler}, {Briois}, {Broiles}, {Burch}, {Calmonte},
  {Cessateur}, {Combi}, {De Keyser}, {Fiethe}, {Galand}, {Gasc}, {Gombosi},
  {Gunell}, {Hansen}, {H{\"a}ssig}, {J{\"a}ckel}, {Korth}, {Le Roy}, {Mall},
  {Mandt}, {Petrinec}, {Raghuram}, {R{\`e}me}, {Rinaldi}, {Rubin}, {S{\'e}mon},
  {Trattner}, {Tzou}, {Vigren}, {Waite}, \& {Wurz}}]{fuselier2015}
{Fuselier}, S.~A., {Altwegg}, K., {Balsiger}, H., {et~al.} 2015, \aap, 583, A2

\bibitem[{{Halpern} {et~al.}(1988){Halpern}, {Miller}, {Okabe}, \&
  {Nottingham}}]{halpern1988}
{Halpern}, J.~B., {Miller}, G.~E., {Okabe}, H., \& {Nottingham}, W. 1988, J.
  Photochem. Photobiol. A, 42, 63

\bibitem[{{Haser}(1957)}]{haser1957}
{Haser}, L. 1957, Bulletin de la Societe Royale des Sciences de Liege, 43, 740

\bibitem[{{Herbst} \& {van Dishoeck}(2009)}]{herbst2009}
{Herbst}, E. \& {van Dishoeck}, E.~F. 2009, \araa, 47, 427

\bibitem[{Hänni {et~al.}(2020)Hänni, Altwegg, Pestoni, Rubin, Schroeder,
  Schuhmann, \& Wampfler}]{haenni2020}
Hänni, N., Altwegg, K., Pestoni, B., {et~al.} 2020, Monthly Notices of the
  Royal Astronomical Society, 498, 2239

\bibitem[{{Huebner} {et~al.}(1992){Huebner}, {Keady}, \& {Lyon}}]{huebner1992}
{Huebner}, W.~F., {Keady}, J.~J., \& {Lyon}, S.~P. 1992, \apss, 195, 1

\bibitem[{{Kanda} {et~al.}(1999){Kanda}, {Nagata}, \& {Ibuki}}]{kanda1999}
{Kanda}, K., {Nagata}, T., \& {Ibuki}, T. 1999, Chemical Physics, 243, 89

\bibitem[{{Kunde} {et~al.}(1981){Kunde}, {Aikin}, {Hanel}, {Jennings},
  {Maguire}, \& {Samuelson}}]{kunde1981}
{Kunde}, V.~G., {Aikin}, A.~C., {Hanel}, R.~A., {et~al.} 1981, \nat, 292, 686

\bibitem[{{Lara} {et~al.}(1996){Lara}, {Lellouch}, {L{\'o}pez-Moreno}, \&
  {Rodrigo}}]{lara1996}
{Lara}, L.~M., {Lellouch}, E., {L{\'o}pez-Moreno}, J.~J., \& {Rodrigo}, R.
  1996, \jgr, 101, 23261

\bibitem[{{L{\"a}uter} {et~al.}(2020){L{\"a}uter}, {Kramer}, {Rubin}, \&
  {Altwegg}}]{laeuter2020}
{L{\"a}uter}, M., {Kramer}, T., {Rubin}, M., \& {Altwegg}, K. 2020, \mnras,
  498, 3995

\bibitem[{{Le Roy} {et~al.}(2015){Le Roy}, {Altwegg}, {Balsiger}, {Berthelier},
  {Bieler}, {Briois}, {Calmonte}, {Combi}, {De Keyser}, {Dhooghe}, {Fiethe},
  {Fuselier}, {Gasc}, {Gombosi}, {H{\"a}ssig}, {J{\"a}ckel}, {Rubin}, \&
  {Tzou}}]{leroy2015}
{Le Roy}, L., {Altwegg}, K., {Balsiger}, H., {et~al.} 2015, \aap, 583, A1

\bibitem[{{Lellouch} {et~al.}(2019){Lellouch}, {Gurwell}, {Moreno}, {Vinatier},
  {Strobel}, {Moullet}, {Butler}, {Lara}, {Hidayat}, \&
  {Villard}}]{lellouch2019}
{Lellouch}, E., {Gurwell}, M.~A., {Moreno}, R., {et~al.} 2019, Nature
  Astronomy, 3, 614

\bibitem[{Loison {et~al.}(2015)Loison, Hébrard, Dobrijevic, Hickson, Caralp,
  Hue, Gronoff, Venot, \& Bénilan}]{loison2015}
Loison, J., Hébrard, E., Dobrijevic, M., {et~al.} 2015, Icarus, 247, 218

\bibitem[{{Magee} {et~al.}(2009){Magee}, {Waite}, {Mandt}, {Westlake}, {Bell},
  \& {Gell}}]{magee2009}
{Magee}, B.~A., {Waite}, J.~H., {Mandt}, K.~E., {et~al.} 2009, \planss, 57,
  1895

\bibitem[{{Maquet}(2015)}]{maquet2015}
{Maquet}, L. 2015, \aap, 579, A78

\bibitem[{{Mattauch} \& {Herzog}(1934)}]{mattauch1934}
{Mattauch}, J. \& {Herzog}, R. 1934, Zeitschrift fur Physik, 89, 786

\bibitem[{{Mumma} \& {Charnley}(2011)}]{mumma2011}
{Mumma}, M.~J. \& {Charnley}, S.~B. 2011, \araa, 49, 471

\bibitem[{{Pandya} {et~al.}(2012){Pandya}, {Shelat}, {Joshipura}, \&
  {Vaishnav}}]{pandya2012}
{Pandya}, S.~H., {Shelat}, F.~A., {Joshipura}, K.~N., \& {Vaishnav}, B.~G.
  2012, International Journal of Mass Spectrometry, 323-324, 28

\bibitem[{{Rubin} {et~al.}(2019){Rubin}, {Altwegg}, {Balsiger}, {Berthelier},
  {Combi}, {De Keyser}, {Drozdovskaya}, {Fiethe}, {Fuselier}, {Gasc},
  {Gombosi}, {H{\"a}nni}, {Hansen}, {Mall}, {R{\`e}me}, {Schroeder},
  {Schuhmann}, {S{\'e}mon}, {Waite}, {Wampfler}, \& {Wurz}}]{rubin2019}
{Rubin}, M., {Altwegg}, K., {Balsiger}, H., {et~al.} 2019, \mnras, 489, 594

\bibitem[{{Schroeder} {et~al.}(2019){Schroeder}, {Altwegg}, {Balsiger},
  {Berthelier}, {De Keyser}, {Fiethe}, {Fuselier}, {Gasc}, {Gombosi}, {Rubin},
  {S{\'e}mon}, {Tzou}, {Wampfler}, \& {Wurz}}]{schroeder2019}
{Schroeder}, I. R.~H.~G., {Altwegg}, K., {Balsiger}, H., {et~al.} 2019, \aap,
  630, A29

\bibitem[{{Steins}(2018)}]{NIST}
{Steins}, S.~E. 2018, { Chemistry WebBook, NIST Standard Reference Database
  Number 69, National Institute of Standards and Technology}

\bibitem[{{Stevenson}(1950)}]{stevenson1950}
{Stevenson}, D.~P. 1950, \jcp, 18, 1347

\bibitem[{{Swings} \& {Haser}(1956)}]{swings1956}
{Swings}, P. \& {Haser}, L. 1956, {Atlas of representative cometary spectra}

\bibitem[{{Teanby} {et~al.}(2009){Teanby}, {Irwin}, {de Kok}, {Jolly},
  {B{\'e}zard}, {Nixon}, \& {Calcutt}}]{teanby2009}
{Teanby}, N.~A., {Irwin}, P.~G.~J., {de Kok}, R., {et~al.} 2009, \icarus, 202,
  620

\bibitem[{{Teanby} {et~al.}(2006){Teanby}, {Irwin}, {de Kok}, {Nixon},
  {Coustenis}, {B{\'e}zard}, {Calcutt}, {Bowles}, {Flasar}, {Fletcher},
  {Howett}, \& {Taylor}}]{teanby2006}
{Teanby}, N.~A., {Irwin}, P.~G.~J., {de Kok}, R., {et~al.} 2006, \icarus, 181,
  243

\bibitem[{{Thelen} {et~al.}(2019){Thelen}, {Nixon}, {Chanover}, {Cordiner},
  {Molter}, {Teanby}, {Irwin}, {Serigano}, \& {Charnley}}]{thelen2019}
{Thelen}, A.~E., {Nixon}, C.~A., {Chanover}, N.~J., {et~al.} 2019, \icarus,
  319, 417

\bibitem[{{van Dishoeck}(2014)}]{vandishoeck2014}
{van Dishoeck}, E.~F. 2014, Faraday Discussions, 168, 9

\bibitem[{{Vastel} {et~al.}(2019){Vastel}, {Loison}, {Wakelam}, \&
  {Lefloch}}]{vastel2019}
{Vastel}, C., {Loison}, J.~C., {Wakelam}, V., \& {Lefloch}, B. 2019, \aap, 625,
  A91

\bibitem[{{Wilson} \& {Atreya}(2004)}]{wilson2004}
{Wilson}, E.~H. \& {Atreya}, S.~K. 2004, Journal of Geophysical Research
  (Planets), 109, E06002

\bibitem[{{Wurz} {et~al.}(2015){Wurz}, {Rubin}, {Altwegg}, {Balsiger},
  {Berthelier}, {Bieler}, {Calmonte}, {De Keyser}, {Fiethe}, {Fuselier},
  {Galli}, {Gasc}, {Gombosi}, {J{\"a}ckel}, {Le Roy}, {Mall}, {R{\`e}me},
  {Tenishev}, \& {Tzou}}]{wurz2015}
{Wurz}, P., {Rubin}, M., {Altwegg}, K., {et~al.} 2015, \aap, 583, A22

\end{thebibliography}

\begin{appendix} 

\section{Results of the correlation analysis}\label{app:1}
The C$_2$N$_2$ signal observed in March 2016, as shown in Fig.~\ref{fig:mar16}, was correlated to the signals of the other species analyzed by \citet{rubin2019}. The results are listed in the following Table~\ref{tab:correl}:

\begin{table}[h!t]
\caption[]{Results of the correlation analysis performed based on data from March 2016.}
\begin{tabular}{ll}
\hline\hline
Species           &    R($c_\mathrm{C_2N_2}$)$^{(a)}$  \\  
\hline
$c_\mathrm{H_2O}$         &    0.39          \\
$c_\mathrm{CO_2}$         &    0.59          \\
$c_\mathrm{CO}$           &    0.65$^{(b)}$  \\
$c_\mathrm{O_2}$          &    0.59          \\
$c_\mathrm{H_2S}$         &    0.84          \\
$c_\mathrm{NH_3}$         &    0.29          \\
$c_\mathrm{CH_4}$         &    0.74          \\
$c_\mathrm{CH_2O}$        &    0.70          \\
$c_\mathrm{C_2H_6}$       &    0.67          \\
$c_\mathrm{CH_3OH}$       &    0.76          \\
$\bm{c_\mathrm{CHN}}$     & \textbf{0.86}    \\
$c_\mathrm{SO_2}$         &    0.35          \\
$c_\mathrm{CH_3CHO}$      &    0.60          \\
$c_\mathrm{OCS}$          &    0.88          \\
$c_\mathrm{HNCO}$         &    0.68          \\
$c_\mathrm{C_3H_8}$       &    0.89          \\
$c_\mathrm{CH_3CN}$       &    0.78          \\
$c_\mathrm{CS_2}$         &    0.81          \\
$c_\mathrm{CH_3NO}$       &    0.59          \\
$c_\mathrm{C_2H_4O_2}$    &    0.84          \\
$c_\mathrm{C_6H_6}$       &    0.78          \\
$c_\mathrm{HC_3N}$        &    0.24          \\
\hline
\end{tabular}
\tablefoot{$^{(a)}$ The resulting Pearson correlation coefficients (R) are arranged according to decreasing bulk abundance relative to water of the isomer considered by \citet{rubin2019} in their Table 2. $^{(b)}$ As $c_\mathrm{CO}$ sometimes is dominated by the CO fragment of CO$_2$, this correlation coefficient may not be representative.}
\label{tab:correl}
\end{table}

\section{Neutral density ratio data for cyanogen}\label{app:2}
For the convenience of the reader, we list the results of our analysis regarding cometary cyanogen  separately, namely the neutral density ratio $n_\mathrm{NCCN}/n_\mathrm{HCN}$ as a function of the Rosetta mission time. These data were shown in Fig.~\ref{fig:fullmiss-hcn} (bottom).

\begin{table*}[h!t]
\caption[]{Neutral density ratio of cyanogen and hydrogen cyanide $n_\mathrm{NCCN}/n_\mathrm{HCN}$ as a function of the Rosetta mission time. $r_\mathrm{h}$ is the heliocentric distance.}
\begin{tabular}{llllll}
\hline\hline
Date          & $r_\mathrm{h}$ [au]  &    $n_\mathrm{NCCN}/n_\mathrm{HCN}$   & Stat. error [\%] & No. of spectra averaged$^{(a)}$ & Detector row$^{(b)}$  \\  \hline
2014-10-17    &    3.16     &    $3.4 \times 10^{-4}$  &    40            &    1       & A    \\
2014-10-24    &    3.12     &    $2.7 \times 10^{-4}$  &    29            &    1       & A    \\
2015-05-06    &    1.70     &    $2.5 \times 10^{-3}$  &    33            &    1       & A    \\
2015-05-07    &    1.69     &    $1.4 \times 10^{-3}$  &    41            &    1       & A    \\
2015-07-30    &    1.25     &    $1.6 \times 10^{-3}$  &    29            &    11      & A+B  \\
2015-11-22    &    1.72     &    $5.0 \times 10^{-4}$  &    44            &    2       & A    \\
2016-03-14    &    2.57     &    $1.2 \times 10^{-4}$  &    22            &    15      & B    \\
2016-03-15    &    2.58     &    $1.4 \times 10^{-4}$  &    19            &    15      & B    \\
2016-03-16    &    2.59     &    $1.2 \times 10^{-4}$  &    20            &    12      & B    \\
2016-07-18    &    3.42     &    $6.2 \times 10^{-4}$  &    25            &    2       & A    \\ \hline
\end{tabular}
\tablefoot{$^{(a)}$ We include this value to indicate the statistical basis of the individual data points. $^{(b)}$ Of the two redundant micro-channel plate/linear electron detector array (MCP/LEDA) rows, usually the one with the higher count rate was used.}
\label{tab:app}
\end{table*}

\end{appendix}

\end{document}